\RequirePackage[T1]{fontenc}
\documentclass[conference,letterpaper]{IEEEtran}
\usepackage{amsmath,amssymb,graphicx,booktabs,cite,url,balance}
\usepackage[hidelinks]{hyperref}
\hypersetup{pdftitle={Capacity Credit of Data-Centre Flexibility: The Role of Workload Scheduling and Usable Battery Energy},pdfauthor={Pengyu Ren; Wei Sun; Jiakai Wu; Yifan Wang},pdfsubject={Author-review draft for IEEE PES GM 2027; revision 16}}
\newcommand{\resultrows}{%
S4 & 2468.5 & 15.32 & 13.54 & 13.53 \\
S8 & 2445.4 & 28.84 & 25.49 & 25.30 \\
E8 & 2480.4 & 8.36 & 7.39 & 7.34 \\
U10 & 2489.4 & 3.08 & 2.72 & 2.72 \\
U15 & 2484.2 & 6.15 & 5.44 & 5.44 \\
B2 & 2483.5 & 6.56 & 5.80 & 5.80 \\
B4 & 2472.2 & 13.12 & 11.60 & 11.60 \\
B8 & 2450.1 & 25.96 & 22.94 & 22.78 \\
UB & 2478.2 & 9.64 & 8.52 & 8.52 \\
}

\newcommand{\EUE}{\mathrm{EUE}}
\newcommand{\CC}{\mathrm{CC}}
\begin{document}
\title{Capacity Credit of Data-Centre Flexibility: The Role\\of Workload Scheduling and Usable Battery Energy}
\author{\IEEEauthorblockN{Pengyu Ren, \textit{Student Member, IEEE}, Wei Sun\textsuperscript{*}, \textit{Member, IEEE},\\
Jiakai Wu, \textit{Student Member, IEEE}, and Yifan Wang, \textit{Student Member, IEEE}}
\IEEEauthorblockA{School of Engineering, University of Edinburgh, Edinburgh, United Kingdom\\
\textsuperscript{*}Corresponding author}}
\maketitle
\begin{abstract}
Data centres that offer the same reduction in grid demand can make different contributions to supply adequacy. We estimate the effective load-carrying capability of workload scheduling and on-site batteries by matching expected unserved energy across paired simulations. Longer task deadlines increase capacity credit by giving operators more choice over when to defer and recover work. A fixed evening commitment shifts more work than deadline-based scheduling but delivers a smaller reduction in expected unserved energy. Protected backup reserves leave uninterruptible power supplies with limited capacity credit, while longer discharge durations raise the credit of dedicated batteries at the same response rating. These findings suggest assessing recovery deadlines and dispatchable energy alongside offered demand reductions.
\end{abstract}

\section{Introduction}
Data centres (DCs) can reduce grid imports by deferring computing tasks or discharging on-site batteries. Scheduling can avoid coincident peaks \cite{liu} while preserving daily service capacity \cite{google}. Flexible operation therefore offers a way to integrate growing demand \cite{esig}, provided that computing and backup obligations remain satisfied.

Workload traces quantify rescheduling costs \cite{zhou}, field demonstrations test power reductions \cite{phoenix}, and job-level forecasts improve demand visibility \cite{mughees}. For capacity planning, however, controllable power is only a starting point: planners need to establish its dependable contribution during system shortages.

Equal reduction ratings need not yield equal adequacy benefits. Deferred work must return, potentially during continued scarcity. Batteries are limited by usable energy, with protected backup reserves unavailable for dispatch. Storage studies show why duration and charging opportunities affect capacity value \cite{konstantelos,mills}. Peak reduction and annual shifted energy alone therefore cannot establish reliability contribution.

Zeng \emph{et al.} \cite{zeng}assessed generation displaced by participating data centres on an IEEE 33-node distribution network. However, the capacity value of different DC services relative to conventional system resources remains less well established. This requires consistent recovery and reserve constraints, chronological dispatch, and attention to risk-estimation precision \cite{rabecq}.

This paper evaluates the capacity value of different forms of data-centre flexibility within the same power system and examines the additional value of coordinated operation. To the best of our knowledge, this is the first study to assess workload flexibility, UPS and dedicated BESS together, including coordinated UPS–BESS operation, under a common connection limit and EUE-based adequacy benchmark. 

By comparing individual and combined services, we examine how workload recovery and protected backup energy shape their dependable capacity contributions, and whether coordination provides additional value beyond standalone operation. ELCC allows these services and conventional system resources to be compared using the same adequacy metric.
\begin{figure*}[t]
\centering\includegraphics[width=\textwidth]{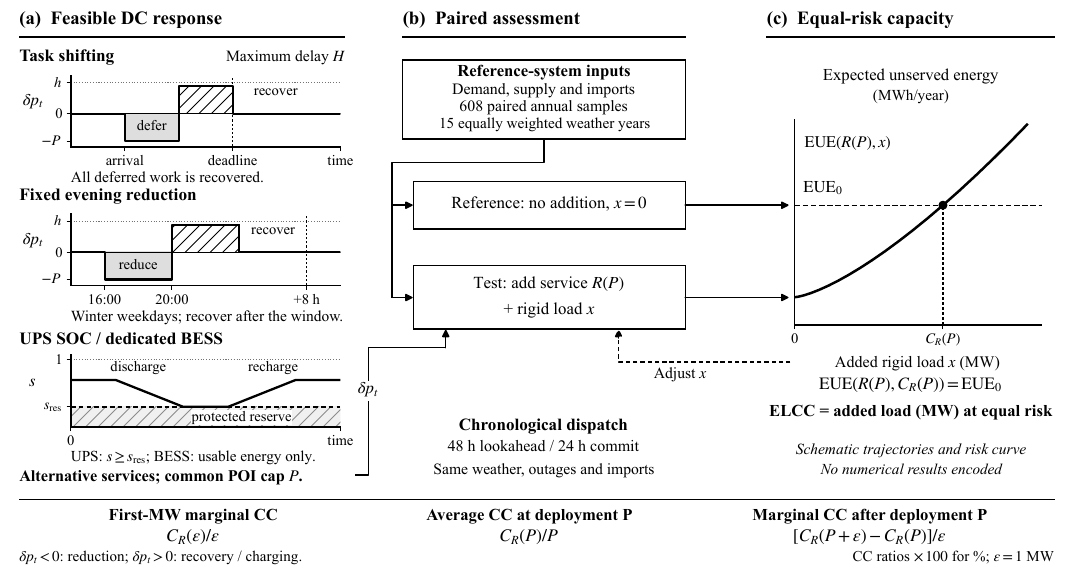}
\caption{Assessment framework: feasible response, paired dispatch and equal-EUE load search.}

\label{fig:workflow}
\end{figure*}

\section{Service Models and Capacity-Credit Method}
Figure~\ref{fig:workflow} summarises the assessment in three stages. 

First, each data-centre service is represented by its feasible change in grid demand: workload shifting must recover deferred tasks within the specified deadline, fixed evening reductions create subsequent recovery obligations, and battery dispatch respects usable-energy limits and protected UPS reserves. These constraints determine when and for how long each service can support the system under a common connection limit. 

Second, chronological dispatch evaluates the reference system and each service configuration using identical weather, outage and import samples, carrying workload backlogs and stored energy between dispatch windows. This paired assessment captures both the immediate reduction in shortages and the effects of subsequent recovery or charging. 

Third, additional inflexible demand is introduced into the system with the evaluated service until EUE returns to the reference level. The resulting demand increment defines ELCC, from which average and marginal capacity credits are calculated.

\subsection{Modeling of compute response at the grid connection}
Deferral reduces dynamic power but retains idle consumption. With reference compute utilisation $u$ and idle-to-full-load power ratio $r$, the affine model \cite{fan} partitions participant mean demand $L$ into dynamic power $B$ and idle power. Deferrable work fraction $f$ sets reduction capability $P$. Recovery headroom $h$ is bounded by unused compute capacity and import allowance $gL$, where $g$ is the import headroom ratio (powers in MW):

\begin{align}
B&=L\frac{(1-r)u}{r+(1-r)u},\qquad P=fB,\nonumber\\
h&=\min\{B(1/u-1),\ gL\},
\label{eq:power}
\end{align}

The envelope uses mean demand; Fraction $f$ concerns computational work here. Scheduling preserves work and deadlines \cite{liu,google}. Let $t$ index intervals of length $\Delta t=1$ h; $a_t=P\Delta t$ and $z_t$ are arriving and completed work (dynamic-energy-equivalent MWh). Cumulative bounds enforce maximum delay $H$ (h) and prevent advance processing; the service bound limits recovery ($\tau$ indexes summation):

\begin{equation}
\sum_{\tau=1}^{t-H}a_\tau\leq\sum_{\tau=1}^{t}z_\tau\leq\sum_{\tau=1}^{t}a_\tau,\quad
0\leq z_t\leq(P+h)\Delta t.
\label{eq:queue}
\end{equation}
Empty sums are zero. Grid-import change is $\delta p_t=z_t/\Delta t-P$: negative values reduce demand. Cohorts carry across dispatch windows and the annual queue closes.

The evening service requires $\delta p_t=-P$ throughout 16:00--20:00 local time on November--March weekdays, including adequate hours. Recovery is due within eight hours after each window, subject to $h$ and year-end closure. Both services conserve work; only the evening deadline is window-based.

\subsection{Modeling of UPS reserve and BESS}
Grid-interactive UPS operation preserves backup energy for data centre operation \cite{eaton}. Let $L_{\rm crit}$ be nominal critical load (MW), $\sigma$ battery state of health, and $\tau_{\rm tot}$ and $\tau_{\rm res}$ total and protected runtimes (h). Subtracting reserve $E_{\rm res}=L_{\rm crit}\tau_{\rm res}$ from healthy UPS energy gives usable energy $E_U$. Dedicated BESS adds usable energy $E_B$ at duration $d_B$ (h):
\begin{equation}
E_U=L_{\rm crit}(\sigma\tau_{\rm tot}-\tau_{\rm res}),\qquad E_B=Pd_B,
\label{eq:energy}
\end{equation}
with $E_U\geq0$. UPS reserve remains fixed during compute reduction; dedicated BESS has no additional backup obligation.

Let $e_t$ be dispatchable stored energy and $E$ its capacity, in grid-deliverable MWh above reserve or cutoff. Charging and discharge powers are $c_t,d_t$ (MW), with efficiencies $\eta_c$ and $\eta_d=1$. Over $T=8760$ hourly intervals, the energy balance, usable bounds and terminal condition are
\begin{align}
e_{t+1}&=e_t+\eta_c c_t\Delta t-d_t\Delta t,\nonumber\\
0&\leq e_t\leq E,\qquad e_0=e_T.
\label{eq:storage}
\end{align}
For UPS, let $E_{\rm max}$ be healthy energy including reserve, $s$ the state-of-charge (SOC) proxy and $s_{\rm res}$ its protected floor. Then
\begin{align}
E_{\rm max}&=E_U+E_{\rm res},\qquad s_{\rm res}=\frac{E_{\rm res}}{E_{\rm max}},\nonumber\\
s&=\frac{e_t+E_{\rm res}}{E_{\rm max}}\geq s_{\rm res}.
\label{eq:soc}
\end{align}
Thus Fig.~\ref{fig:workflow}(a) shows total stored energy normalised by healthy capacity. Empty BESS energy instead denotes technical cutoff, not zero cell SOC. Battery import change is $c_t-d_t$; combined assets share discharge limit $P$ and charge limit $\min(P,gL)$.

\begin{table*}[t]
\caption{Case Parameters and Evaluated Configurations}
\label{tab:parameters}
\centering\footnotesize\setlength{\tabcolsep}{3pt}
\begin{minipage}[t]{0.49\textwidth}
\centering
\begin{tabular}[t]{p{1.88in}cr}
\toprule Full parameter name & Symbol & Value\\\midrule
Participating demand fraction & --- & 30\%\\
Deferrable work fraction & $f$ & 0.20\\
Reference compute utilisation & $u$ & 0.80\\
Idle-to-full-load power ratio & $r$ & 0.50\\
Incremental import headroom ratio & $g$ & 0.10\\
Nominal critical-load proxy & $L_{\rm crit}$ & $L$\\
UPS battery state of health & $\sigma$ & 1.00\\
Protected UPS backup runtime & $\tau_{\rm res}$ & 5 min\\
Charging / discharge efficiency & $\eta_c/\eta_d$ & 0.85 / 1\\
Initial / final dispatchable energy & $e_0,e_T$ & $0.5E$\\
Extra dedicated BESS reserve & --- & 0 MWh\\\midrule
\multicolumn{3}{l}{Derived reference ratings from (\ref{eq:power})}\\
Participant reference mean power & $L$ & 1272.705 MW\\
Dynamic power component & $B$ & 565.647 MW\\
POI import-reduction capability & $P$ & 113.129 MW\\
Compute recovery headroom & $h$ & 127.271 MW\\
\bottomrule
\end{tabular}
\end{minipage}\hfill
\begin{minipage}[t]{0.49\textwidth}
\centering
\begin{tabular}[t]{lp{2.82in}}
\toprule Case & Technology and distinguishing setting\\\midrule
S4 & Deadline-constrained compute response: $H=4$ h\\
S8 & Deadline-constrained compute response: $H=8$ h\\
E8 & Fixed evening compute reduction; 8 h recovery after the window\\
U10 & Uninterruptible power supply, total runtime $\tau_{\rm tot}=10$ min\\
U15 & Uninterruptible power supply, total runtime $\tau_{\rm tot}=15$ min\\
B2 & Battery energy storage system: $d_B=2$ h\\
B4 & Battery energy storage system: $d_B=4$ h\\
B8 & Battery energy storage system: $d_B=8$ h\\
UB & U10 and B2 together, sharing one POI limit $P$\\
\bottomrule
\end{tabular}
\par\smallskip\raggedright
E8 reduces $P$ throughout 16:00--20:00, November--March weekdays (UK local time), including bank holidays. Other services are available every hour subject to their constraints.
\par\smallskip
UPS minutes are rated at $L_{\rm crit}=11.25P$; BESS hours are rated at $P$. Thus U10/U15 have 0.9375/1.875 h of usable energy per POI MW.
\end{minipage}
\end{table*}

\subsection{Chronological risk assessment}
For weather year $w$, draw $j$ and hour $t$, let $G$ denote non-storage supply, $D$ baseline demand, $b^-,b^+$ storage discharge and charging, $\ell\geq0$ unserved power, $v$ spill and $x$ added rigid demand. All are in MW; $\delta p$ is the service import change:

\begin{equation}
G_{wjt}+b^-_{wjt}+\ell_{wjt}
=D_{wt}+x+\delta p_{wjt}+b^+_{wjt}+v_{wjt}.
\label{eq:balance}
\end{equation}
Hourly linear programmes minimise unserved energy over 48 hours and commit 24, carrying states forward. A secondary objective favours stored energy and earlier completion at equal EUE. 

For added resource configuration $R$ and rigid load $x$, the EUE estimator averages annual unserved energy over $W$ equally weighted weather years and $n_w$ draws in year $w$:
\begin{equation}
\widehat{\EUE}(R,x)=\frac{1}{W}\sum_{w=1}^{W}\frac{1}{n_w}
\sum_{j=1}^{n_w}\sum_{t=1}^{T}\ell_{wjt}(R,x)\Delta t.
\label{eq:eue}
\end{equation}
Alternatives share samples; weather weights remain equal despite unequal $n_w$.

\subsection{Capacity credit, ELCC and marginal contribution}
Capacity credit (CC) quantifies adequacy contribution, here normalised by resource MW. Write $R(P)$ for a service of capability $P$. Following Garver \cite{garver}, its ELCC $C_R(P)$ is additional rigid load (MW) carried at reference risk $\EUE_0$, without service or added load in the reference:
\begin{equation}
\widehat{\EUE}(R(P),C_R(P))=\widehat{\EUE}(0,0)=\EUE_0.
\label{eq:elcc}
\end{equation}
Marginal ELCC measures the local contribution per added MW, consistent with Bothwell and Hobbs' EUE-based capacity analysis \cite{bothwell}. For rating increment $\varepsilon=1$ MW,
\begin{equation}
\frac{dC_R}{dP}\approx\frac{C_R(P+\varepsilon)-C_R(P)}{\varepsilon}.
\label{eq:marginal}
\end{equation}
The numerator is incremental ELCC in MW; the ratio has units MW/MW. First-MW marginal, average and post-deployment marginal capacity-credit ratios are, respectively,

\begin{align}
\CC_{\mathrm{m},R}(0)&=C_R(\varepsilon)/\varepsilon,\nonumber\\
\overline{\CC}_R(P)&=C_R(P)/P,\nonumber\\
\CC_{\mathrm{m},R}(P)&=[C_R(P+\varepsilon)-C_R(P)]/\varepsilon.
\label{eq:credits}
\end{align}
Multiplication by 100 gives CC (\%). Cross-technology plots use $\CC_{\mathrm{m},R}(0)$: zero means no evaluated addition to the reference system, which may already contain that technology.

Increasing $P$ scales work and site capabilities without raising baseline demand; $x$ remains inflexible. System additions scale class availability profiles without independent new-unit outages. ELCC uses signed brackets no wider than 0.002 MW, distinct from sampling uncertainty.

\begin{figure*}[t]
\centering\includegraphics[width=\textwidth]{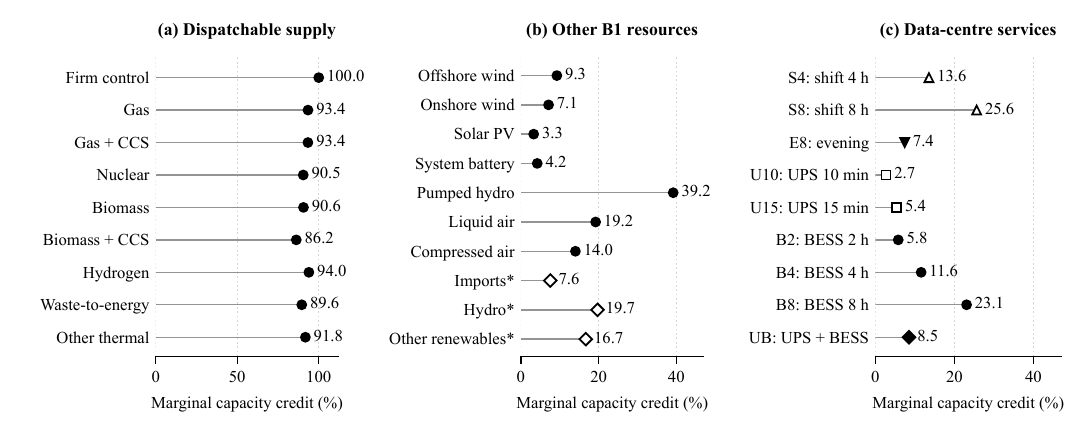}
\caption{Marginal capacity credit at the reference system (B1): independent 1 MW additions, 608 paired samples. Panels (b,c) share a scale; (a) uses a wider scale. Open diamonds (*) mark import or flat-output proxies; DC symbols distinguish computing, UPS, BESS and combined storage. CCS: carbon capture and storage; PV: photovoltaics. Brackets span at most 0.20 percentage points. Denominators: added system-resource MW or DC POI-reduction MW.}
\label{fig:common}
\end{figure*}

\begin{figure}[t]
\centering\includegraphics[width=\columnwidth]{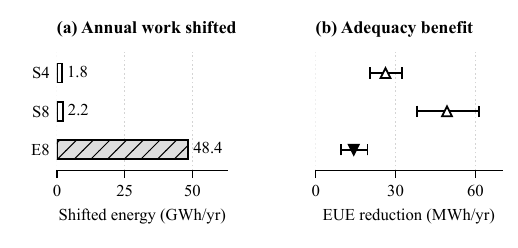}
\caption{Workload services at $P=113.129$ MW. (a) Annual shifted work in its dynamic-energy equivalent; hatching denotes mandatory evening reduction. (b) EUE reduction with 95\% paired bootstrap intervals from 4000 resamples within weather years. Intervals condition on the sampled weather years and model parameters.}
\label{fig:rebound}
\end{figure}

\begin{figure}[t]
\centering\includegraphics[width=\columnwidth]{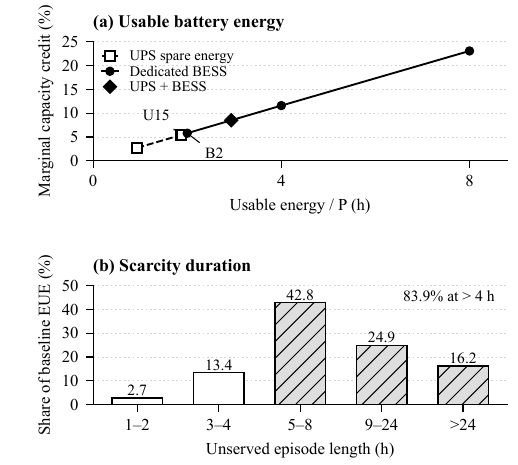}
\caption{(a) Marginal capacity credit at B1 versus usable energy per POI MW; lines join tested sizes and UPS protects five backup minutes. (b) EUE-weighted episode durations: $10^{-5}$ MW threshold, year-boundary termination, hatching beyond four hours. Shares describe energy, not event counts.}
\label{fig:duration}
\end{figure}

\section{Case Study, Results and Interpretation}
\subsection{Planning relevance and system design}
The case study considers Great Britain (GB), where the National Energy System Operator (NESO) projects data-centre demand of 20–41 TWh in 2035 [15]. The 2026 Data Centre Covenant \cite{covenant} and NESO's feasibility study \cite{nesodc} motivate assessing its flexibility.

Reference system B1 uses 2035 Holistic Transition data \cite{fes}, PyPSA-GB's scenario data structure \cite{pypsa} and separate copperplate dispatch. NESO's 2035/36 Portfolio 1 supplies four totals: gas 27.2 GW, gas with carbon capture and storage (CCS) 3.9 GW, nuclear 5.8 GW, and biomass including CCS 2.8 GW \cite{nesora}.

Fifteen weather years (2010--2024) vary wind, solar and weather-sensitive demand over 8760 hours. DC demand uses UK Power Networks' (UKPN) high-percentile planning profile from peak-normalised data for at least ten sites per category \cite{ukpn}. Half-hours are averaged to hours and scaled to 37.163 TWh/year without another power-usage-effectiveness multiplier.

The 608 annual samples follow a 336-hour warmup, giving $\EUE_0=2494.703$ MWh/year with conditional Monte Carlo standard error 400.091 MWh/year. Background storage has a free annual terminal state with an energy-retention preference; added DC batteries restore initial energy. This distinction remains in cross-resource comparisons.

Table~\ref{tab:parameters} defines the cases. Traces inform utilisation \cite{zhou}; scheduling studies support mechanisms rather than selected fractions or deadlines \cite{liu,google,phoenix}. UPS runtimes are design choices \cite{eaton}; BESS duration and efficiency follow the 2024 Annual Technology Baseline \cite{atb}.

\subsection{Data-centre flexibility within the resource mix}
At the first-1-MW comparison point, marginal capacity credit varies substantially across DC services, from 2.7\% for U10 to 25.6\% for S8 (Fig.~\ref{fig:common}). Deadline-based computing and four/eight-hour BESS exceed wind, solar and system batteries, but remain below pumped hydro and thermal supply. UPS-only and fixed-evening services lie lower. Credit is per offered POI-reduction MW for DC services and per added MW for system resources.

Eight-hour deferral earns nearly twice the credit of four-hour deferral: extra recovery time helps avoid returning work during shortages. Thermal capacity can sustain deficits under unconstrained fuel, whereas work requires recovery and batteries have finite energy. Storage duration, availability and terminal conditions also differ across resources.

\subsection{Response energy versus adequacy benefit}
\begin{table}[t]
\caption{DC Deployment Results at $P=113.129$ MW. Marginal CC Uses a Further 1 MW Addition.}
\label{tab:results}
\centering\footnotesize\setlength{\tabcolsep}{3pt}
\begin{tabular}{lrrrr}
\toprule & EUE & ELCC & Average CC & Marginal CC\\
Case & (MWh/yr) & (MW) & (\%) & (\%)\\\midrule
\resultrows
\bottomrule
\end{tabular}
\end{table}
The evening service shifts over twenty times the work of eight-hour deferral but delivers less than one-third of its EUE reduction (Fig.~\ref{fig:rebound}). Mandatory reductions during adequate evenings create recovery obligations without avoiding unserved energy. Optional deferral instead targets shortages and clears its backlog using subsequent compute and import headroom.

Eight-hour deferral has the largest deployment ELCC (Table~\ref{tab:results}); similar average and marginal credits indicate little change over the tested increment. 

\subsection{Capacity support while preserving backup obligations}
UPS dispatch supports adequacy using energy above its protected floor; dedicated BESS adds usable energy without drawing on that reserve. Retaining five backup minutes, extending UPS runtime from ten to fifteen minutes doubles usable energy and nearly doubles marginal credit (Fig.~\ref{fig:duration}(a)).

UPS-only credit remains modest; longer-duration BESS provides greater support at unchanged POI rating. Combined UPS--BESS preserves the backup obligation and connection limit.

Over four-fifths of B1 EUE occurs in shortages exceeding four hours (Fig.~\ref{fig:duration}(b)). Yet most residual U10 EUE occurs in hours ending above reserve: dispatch favours holding energy at equal horizon EUE. Low credit therefore cannot be attributed solely to battery depletion; scheduling also matters.

\section{Conclusion}
At the evaluated deployment scale, eight-hour workload deferral achieves the highest ELCC of 28.84 MW, compared with 8.36 MW for fixed evening reduction. UPS capacity credit is constrained by protected backup energy, while dedicated BESS gains capacity value with longer discharge duration. These results show that offered MW alone is insufficient to characterise the adequacy contribution of DC flexibility; recovery constraints and usable energy must also be considered.

\end{document}